\documentclass[12pt,titlepage,acknowledgements,a4paper]{article}
\usepackage{amsmath}
\usepackage{amsfonts}
\newcommand{\ba}{\bar{a}}
\newcommand{\bb}{\bar{b}}
\newcommand{\bc}{\bar{c}}
\newcommand{\bd}{\bar{d}}

\newcommand{\bg}{\bar{g}}

\newcommand{\bz}{\bar{Z}}
\newcommand{\bp}{\bar{\psi}}
\newcommand{\e}{\epsilon}
\newcommand{\be}{\bar{\epsilon}}

\newcommand{\nn}{\nonumber}
\newcommand{\G}{\Gamma}
\newcommand{\Gt}{\tilde{\Gamma}}
\newcommand{\tr}{\textrm{tr}}
\newcommand{\Tr}{\textrm{Tr}}

\makeatletter
\@addtoreset{equation}{section}
\makeatother

\newcounter{multieqs}

\begin{document}

\begin{flushright}
%arXiv:YYMM.NNNN\\
QMUL-PH-09-05
\end{flushright}

\vspace{20pt}

\begin{center}

{\Large \bf  $\mathcal{N} = 6$ Membrane Worldvolume Superalgebra}\\
\vspace{33pt}

{\bf {\mbox{Andrew M. Low}}}%
\footnote{{{\tt a.m.low}{\tt @qmul.ac.uk }}}

{\em Centre for Research in String Theory \\ Department of Physics\\
Queen Mary, University of
London\\
Mile End Road, London, E1 4NS\\
United Kingdom}
\vspace{40pt}

{\bf Abstract}

\end{center}

\noindent
In arXiv:0806.0363 the worldvolume superalgebra of the $\mathcal{N}=8$ Bagger-Lambert theory was calculated. In this paper we derive the general form for the worldvolume superalgebra of the $\mathcal{N}=6$ Bagger-Lambert theory. For a particular choice of three-algebra we derive the superalgebra of the ABJM theory. We interpret the associated central charges in terms of BPS brane configurations. In particular we find the central charge corresponding to the energy bound of the BPS fuzzy-funnel configuration of the ABJM theory. We also derive general expressions for the BPS equations of the $\mathcal{N}=6$ Bagger-Lambert theory.

\vspace{0.5cm}

\setcounter{page}{0}
\thispagestyle{empty}
\newpage

%%%%%%%%%%%%%%%%%%%%%%%%%%%%%%%%%%%%%%%%%%%%%%%%%%%%%

\setcounter{footnote}{0}

\section{Introduction}
Recently there has been much effort directed towards the study of M$2$-branes (for a review see \cite{Berman2}). Particular interest has surrounded the attempt to formulate a Lagrangian that is capable of describing the low energy dynamics of multiple M$2$-branes. This work began with the efforts of Bagger and Lambert \cite{BL1}-\cite{BL4}(see also Gustavson \cite{GUS}). The original Bagger-Lambert theory is a 3-dimensional $\mathcal{N} = 8$ supersymmetric field theory, based on a novel algebraic structure; the so-called 3-algebra. A $3$-algebra is a vector space with basis $T^a$,  which satisfies a triple product. In \cite{BL2}, the triple-product was required to satisfy two conditions. The first was that it satisfied the fundamental identity (which can be expressed as a condition on the structure constants). The second was that the structure constants$f^{abc}_{\phantom{1}\phantom{1}\phantom{1} d}$ be real and antisymmetric in $a,b,c$. Bagger and Lambert were able to construct a  Chern-Simons Lagrangian for their theory by defining a trace form $h^{ab}$ which acts on the $3$-algebra.
For $h^{ab}$ and $f^{abc}_{\phantom{1}\phantom{1}\phantom{1} d}$ real, gauge invariance implies that $f^{abcd} = f^{abc}_{\phantom{1}\phantom{1}\phantom{1} e} h^{ed}$ is totally antisymmetric. When the metric is positive definite, it has been shown \cite{Nagy}-\cite{Papa} that there is essentially one unique example in which $f^{abcd} \propto \varepsilon^{abcd}$. In \cite{LandT} and \cite{Mukhi} it was shown that this theory describes two M$2$-branes in an ${\mathbb{R}}^8 / {\mathbb{Z}}_2$ orbifold background. It is possible to consider the case of a Lorentzian signature metric and this has been done in for example \cite{Gomis}-\cite{Ez}, however the status of this theory is still somewhat unclear (see also \cite{Mukhi2}). For related work, including higher derivative corrections to BLG theory, see for example \cite{Dan}-\cite{Power}.
\\
\\
Another possibility is to look for theories with fewer supersymmetries. In \cite{Lee1} the most general $\mathcal{N}=4$ super-conformal Chern-Simons theory was constructed. In \cite{aha}, ABJM constructed an $\mathcal{N} = 6$ Chern-Simons theory with $ U(N)\times U(N) $ gauge group and $SO(6)$ $R$-symmetry, and claimed that the theory describes N M2-branes in a ${\mathbb{C}}^4 / {\mathbb{Z}}_k$ orbifold background. In the limit in which the number of branes, N, and the Chern-Simons level $k$ are large, with $\lambda = N/k$ fixed, the theory admits a dual geometric description given by $AdS4 \times \mathbb{C}{\mathbb{P}}^3$. The action for this theory was derived from a superspace perspective in \cite{Benna}, and the supersymmery of the action was shown explicitly in \cite{Band}. Motivated by the work of ABJM and \cite{Lee2}, Bagger and Lambert derived the general form for a three-dimensional scale-invariant field theory with $\mathcal{N} = 6$ supersymetry, $SU(4)$ $R$-symmetry and a $U(1)$ global symmetry \cite{BL3}. This was achieved by relaxing the constraint on the structure constants. They showed that for a specific class of 3-algebra one recovers the $\mathcal{N} = 6$ theory of \cite{aha}.
\\
\\
The original motivation of Bagger and Lambert was to write down a theory capable of reproducing the Basu-Harvey equation as a BPS equation. The energy bound corresponding to this particular BPS configuration should appear in the superalgebra of the theory as a central charge term. This was found to be the case for the $\mathcal{N}=8$ Bagger-Lambert theory in \cite{Pass} (See also \cite{Kazu} for space-time superalgebra). For work on BPS configurations and M-brane bound states of the $\mathcal{N}=8$ theory see \cite{IJeon}-\cite{JKIM}. One would expect similar results for the $\mathcal{N}=6$ theory of ABJM. It was shown in \cite{Gomis3, Terashima, KHHL} that the ABJM theory admits fuzzy $S^3$ and fuzzy-funnel BPS solutions. For an alternative derivation of the Basu-Harvey equation see \cite{Douglas}.
\\
\\
In this paper we will compute the extended worldvolume superalgebra for the generalised $\mathcal{N} = 6$ Bagger-Lambert theory. For a particular choice of $3$-algebra we are able to derive the worldvolume superalgebra of the ABJM theory with central charge terms. We find that the central charge corresponding to the half-BPS fuzzy funnel configuration of the ABJM theory appears as a diagonal element of the superalgebra. We also derive the general BPS equations of the $\mathcal{N}=6$ Bagger-Lambert theory. 
\\
\\
The paper is organised as follows. In the next section we will briefly review the $\mathcal{N}=6$ Bagger-Lambert construction and its relation to the ABJM theory. In section 3 we will explicitly calculate the superalgebra associated with the $\mathcal{N}=6$ Bagger-Lambert theory. In section 4 we will derive the ABJM superalgebra using the result of section 3. In section 5 we will derive the general BPS equations of the Bagger-Lambert theory. Finally we will provide some concluding remarks in section 6. In the Appendix we outline our conventions and include calculational details.

\section{$\mathcal{N} = 6$ Bagger-Lambert Theory} \label{section1}
In this section we will briefly review the $\mathcal{N}= 6$ construction of \cite{BL3} and its relation to the ABJM theory of \cite{aha}. A $3$-algebra is defined as a vector space with basis $T^a$ , $a = 1 \ldots  N$, endowed with a triple product
\begin{equation}
[T^a , T^b ; T^{\bc}] = f^{ab \bc}_{\phantom{1} \phantom{1} \phantom{1} d} T^d .
\end{equation} 
Here we follow \cite{BL3} and take the $3$-algebra to be a complex vector space and only demand that the triple product be antisymmetric in the first two indices. Furthermore the $f^{ab \bc}_{\phantom{1} \phantom{1} \phantom{1} d}$ are required to satisfy the following fundamental identity,
\begin{equation}
f^{ef\bg}_{\phantom{1}\phantom{1}\phantom{1} b} f^{cb \ba}_{\phantom{1}\phantom{1}\phantom{1} d} + f^{fe\ba}_{\phantom{1}\phantom{1}\phantom{1} b} f^{cb\bg}_{\phantom{1}\phantom{1}\phantom{1} d} + f^{*\bg \ba f}_{\phantom{1}\phantom{1}\phantom{1} \phantom{1}\bb} f^{ce \bb}_{\phantom{1}\phantom{1}\phantom{1} d} + f^{*\ba \bg e}_{\phantom{1}\phantom{1}\phantom{1} \phantom{1}\bb} f^{cf \bb}_{\phantom{1}\phantom{1}\phantom{1} d} = 0 .
\end{equation}
In order to construct a Lagrangian it is necessary to define a trace form on the $3$-algebra which provides a notion of an inner product, namely
\begin{equation}
h^{\ba b} = \Tr (T^{\ba} , T^b ). \label{gauge}
\end{equation}
Gauge-invariance of the Lagrangian requires that the metric defined by \eqref{gauge} be gauge invariant. In order for this to be true it can be shown \cite{BL3} that the structure constants $f^{ab \bc \bd}$ must satisfy
\begin{equation}
f^{ab \bc \bd} = f^{* \bc \bd a b}.
\end{equation}
In other words complex conjugation acts on $f^{ab \bc \bd}$ as
\begin{equation}
(f^{ab\bc\bd})^* = f^{*\ba \bb c d} = f^{cd \ba \bb}. \label{com}
\end{equation}
Given this information Bagger and Lambert were able to construct the following Lagrangian
\begin{align}
\mathcal{L} =& -D^\mu \bz^a_A D_\mu Z^A_a - i \bp^{Aa} \gamma^\mu D_\mu \psi_{Aa} - V + \mathcal{L}_{CS} \nonumber\\
& -i f^{ab \bc \bd} \bp^A_{\bd} \psi_{Aa} Z^B_b \bz_{B \bc} + 2i f^{ab \bc \bd} \bp^A_{\bd} \psi_{Ba} Z^B_b \bz_{A \bc} \label{zop} \\
& + \frac{i}{2} \varepsilon_{ABCD} f^{ab \bc \bd} \bp^A_{\bd} \psi^B_{\bc} Z^C_a Z^D_b - \frac{i}{2} \varepsilon^{ABCD} f^{cd \ba \bb} \bp_{Ac} \psi_{Bd} \bz_{C \ba} \bz_{D \bb} \nonumber, 
\end{align}
with the potential given by
\begin{equation}
V = \frac{2}{3} \Upsilon^{CD}_{Bd} {\bar{\Upsilon}}^{Bd}_{CD} \label{potential}
\end{equation}
where
\begin{align}
\Upsilon^{CD}_{Bd} &= f^{ab\bc}_{\phantom{1}\phantom{1}\phantom{1} d} Z^C_a Z^D_b \bz_{B \bc} - \frac{1}{2} \delta^C_B f^{ab\bc}_{\phantom{1}\phantom{1}\phantom{1} d} Z^E_a Z^D_b \bz_{E \bc} + \frac{1}{2} \delta^D_B f^{ab\bc}_{\phantom{1}\phantom{1}\phantom{1} d} Z^E_a Z^C_b \bz_{E\bc}
\end{align}
and the twisted Chern-Simons term $\mathcal{L}_{CS}$ is given by
\begin{equation}
\mathcal{L}_{CS} = \frac{1}{2} \varepsilon^{\mu \nu \lambda} \left( f^{ab \bc \bd} A_{\mu \bc b} \partial_\nu A_{\lambda \bd a} + \frac{2}{3} f^{ac\bd}_{\phantom{1}\phantom{1}\phantom{1} g} f^{ge \bar{f} \bb} A_{\mu \bb a} A_{\nu \bd c} A_{\lambda \bar{f} e} \right).
\end{equation}
The $Z^A_a$ are four complex $3$-algebra valued scalar fields with $A=1,2,3,4,$. Their complex conjugates are written as $\bz_{A \ba} = (Z^A_a)^*$. We write the fermions as $\psi_{Aa}$ and their complex conjugates as $\psi^A_{\ba} = (\psi_{Aa})^*$. Note that the act of complex conjugation raises and lowers the $A$ index and interchanges $a \leftrightarrow \ba$. When the $A$ index is raised it means that the corresponding field transforms in the $4$ of $SU(4)$ and a lowered index field transforms in the $\bar{4}$. The covariant derivative is defined by $D_\mu Z^A_d = \partial_\mu Z^A_d - {\tilde{A}}^{\phantom{1}c}_{\mu \phantom{1} d} Z^A_c$. It follows that $D_\mu \bz_{A \bd} = \partial_\mu \bz_{A \bd} - {\tilde{A}}^{* \bc}_{\mu \phantom{1} \bd} \bz_{A \bc}$. Supersymmetry requires that $D_\mu \psi^A_{\bd} = \partial_\mu \psi^A_{\bd} - {\tilde{A}}^{* \bc}_{\mu \phantom{1} \bd} \psi^A_{\bc}$ and $D_\mu \psi_{Ad} - {\tilde{A}}^{\phantom{1}c}_{\mu \phantom{1} d} \psi_{Ac}$. The gauge field kinetic term is of Chern-Simons type and thus does not lead to propagating degrees of freedom. The above Lagrangian is invariant under the following supersymmetry transformations 
\begin{align}
\delta Z^A_a &= i {\bar{\epsilon}}^{AB} \psi_{Ba} \nonumber \\
\delta \psi_{Aa} &= - \gamma^\mu D_\mu Z^B_a \e_{AB} - f^{db \bar{c}}_{\phantom{1}\phantom{1}\phantom{1}a} Z^C_d Z^B_b \bz_{C \bar{c}} \e_{AB} + f^{db \bar{c}}_{\phantom{1}\phantom{1}\phantom{1}a} Z^C_d Z^D_b \bz_{A \bar{c}} \e_{CD} \label{susy1}\\
\delta {\tilde{A}}^{\phantom{1}c}_{\mu \phantom{1} d} &= -i \be_{AB} \gamma_\mu Z^A_a \psi^B_{\bb} f^{ca\bb}_{\phantom{1}\phantom{1}\phantom{1}d} + i \be^{AB} \gamma_\mu \bz_{A \bb} \psi_{Ba} f^{ca\bb}_{\phantom{1}\phantom{1}\phantom{1}d} \nonumber 
\end{align}
up to a surface term (See Appendix \ref{surface}). The supersymmetry parameters $\e_{AB}$ are in the $6$ of $SU(4)$. They satisfy the reality condition $\epsilon^{AB} = \frac{1}{2} \varepsilon^{ABCD} \epsilon_{CD}$. The supersymmetry algebra closes into a translation plus a gauge transformation. As shown in \cite{BL3}, the $f^{ab\bc \bd}$ generate the Lie algebra $\mathcal{G}$ of gauge transformations. In particular if the Lie algebra $\mathcal{G}$ is of the form
\begin{equation}
\mathcal{G} = \otimes_\lambda \mathcal{G}_\lambda
\end{equation}
where $\mathcal{G}_\lambda$ are commuting subalgebras of $\mathcal{G}$, then 
\begin{equation}
f^{ab \bc \bd} = \sum_\lambda \omega_\lambda \sum_\alpha (t^\alpha_\lambda)^{a \bd} (t^\alpha_\lambda)^{b \bc} , 
\end{equation}
where the $t^\alpha_\lambda$ span a $u(N)$ representation of the generators of $\mathcal{G}_\lambda$ and the $\omega_\lambda$ are arbitrary constants. This form of $f^{ab \bc \bd}$ allows one to rewrite the Lagrangian \eqref{zop} as
\begin{align}
\mathcal{L} = &-\Tr (D^\mu \bz_A , D_\mu Z^A) - i \Tr (\bp^A , \gamma^\mu D_\mu \psi_A) - V + \mathcal{L}_{CS}  \nn\\
&- i \Tr (\bp^A , [\psi_A , Z^B ; \bz_B ]) + 2i \Tr (\bp^A , [\psi_B , Z^B ; \bz_A ])\label{zip} \\ 
&+ \frac{i}{2} \varepsilon_{ABCD} \Tr (\bp^A , [Z^C , Z^D ; \psi^B ]) - \frac{i}{2} \varepsilon^{ABCD} \Tr (\bz_D , [\bp_A , \psi_B ; \bz_C ]), \nn
\end{align}
where now
\begin{align}
V = \frac{2}{3} \Tr (\Upsilon^{CD}_B , {\bar{\Upsilon}}^B_{CD}),
\end{align}
with
\begin{align}
\Upsilon^{CD}_B &= [Z^C, Z^D ; \bz_B ] - \frac{1}{2} \delta^C_B [Z^E, Z^D ; \bz_E] + \frac{1}{2} \delta^D_B [Z^E , Z^C ; \bz_E ].
\end{align}
The equivalence of \eqref{zip} and \eqref{zop} can be verified by expanding the fields $Z^A , \psi_A$ in terms of the generators $T^a$ and defining the trace form as in \eqref{gauge}. For example
\begin{align}
\Tr (\bp^A , [\psi_A , Z^B ; \bz_B ]) &= \Tr ( \bp^A_{\bd} T^{\bd} , [\psi_{A a} T^a , Z^B_b T^b ; \bz_{B \bc} T^{\bc}] ) \nn\\
&= \bp^A_{\bd} \psi_{A a} Z^B_b \bz_{B \bc} \Tr (T^{\bd} , [T^a , T^b , T^{\bc}]) \nn \\
&= \bp^A_{\bd} \psi_{A a} Z^B_b \bz_{B \bc} f^{ab \bc \bd}.
\end{align}
In \cite{BL3} it was shown that for a particular choice of triple product one is able to recover the $N=6$ Lagrangian of ABJM written in component form \cite{aha, Benna}. Given two complex vector spaces $V_1$ and $V_2$ of dimension $N_1$ and $N_2$ respectively one may consider the vector space $\mathcal{A}$ of linear maps $X: V_1 \rightarrow V_2$. A triple product may be defined on $\mathcal{A}$ as
\begin{equation}
[X, Y ; Z] = \lambda (X Z^\dagger Y - Y Z^\dagger X) \label{bracket}
\end{equation}
where $\dagger$ denotes the transpose conjugate and $\lambda$ is an arbitrary constant. The inner product acting on this space may be written as
\begin{equation}
\Tr( X, Y) = \tr (X^\dagger Y). \label{trace}
\end{equation}
With this choice of 3-algebra, the Lagrangian \eqref{zip} takes the form of the ABJM theory Lagrangian presented in \cite{Benna}. In the next section we will calculate the superalgebra for the $\mathcal{N}=6$ Bagger-Lambert theory and express the central charges in terms of 3-brackets. We can then make use of \eqref{bracket} and \eqref{trace} to derive the ABJM central charges.
     
\section{$\mathcal{N}=6$ Bagger-Lambert Superalgebra} 
In this section we will calculate the superalgebra associated with the general $\mathcal{N}=6$ Bagger-Lambert Lagrangian. We will follow the method outlined in \cite{Pass}. Given the invariance of the Lagrangian under the supersymmetry variations \eqref{susy1}, Noether's theorem implies the existence of a conserved supercurrent $J^\mu$. The  supercharge is  the spatial integral over the worldvolume coordinates of the zeroth component of the supercurrent. Since we know that the supercharge is the generator of supersymmetry transformations and that the infinitesimal variation of an anticommuting field is given by $\delta \Phi \propto \{ Q, \Phi \} $ we can write 
\begin{equation}
\int d^2 \sigma \delta J^I_{0 \beta} = \{Q^I_\alpha ,Q^J_\beta\}{\bar{\epsilon}}^\alpha_J \label{qerd}
\end{equation}
In order to make use of \eqref{qerd} in the form presented, we will have to re-write the parameters $\epsilon_{AB}$ appearing in the Bagger-Lambert theory in terms of a basis of  $4 \times 4$ gamma matrices,
\begin{equation}
\epsilon_{AB} = \varepsilon^I .(\G^I_{AB}),
\end{equation} 
with $I= 1, \ldots 6$. The $\varepsilon^I$ are carrying a suppressed worldsheet spinor index and represent the $\mathcal{N} = 6$ SUSY generators. The gamma matrices are antisymmetric $(\G^I_{AB} = -\G^I_{BA})$ and satisfy the reality condition 
\begin{equation}
\Gt^{IAB} = \frac{1}{2} \varepsilon^{ABCD} \G^I_{CD} = - (\G^I_{AB})^*. \label{up}
\end{equation}
Furthermore they satisfy\footnote{One explicit realisation in terms of Pauli matrices \cite{Terashima} is given by $\G^1 = \sigma_2 \otimes 1_2, \G^2 = -i \sigma_2 \otimes \sigma_3 , \G^3 = i \sigma_2 \otimes \sigma_1 , \G^4 = - \sigma_1 \otimes \sigma_2 , \G^5 = \sigma_3 \otimes \sigma_2 , \G^6 = -i 1_2 \otimes \sigma_2.$}
\begin{equation}
\G^I_{AB} \Gt^{JBC} + \G^J_{AB} \Gt^{IBC} = 2 \delta^{IJ} \delta^C_A.
\end{equation}
We note that the $4 \times 4$ matrices $\G^I$ act on a different vector space to the $2 \times 2$ matrices $\gamma^\mu$ which are defined as world volume gamma matrices. The supercurrent can be calculated by the usual Noether method.  In general one has
\begin{equation}
J^\mu = \frac{\partial \mathcal{L}}{\partial (\partial_\mu \varphi )} \delta \varphi - V^\mu 
\end{equation}
where $\varphi$ represents all the fields appearing in the Lagrangian and $\delta \mathcal{L} = \partial_\mu V^\mu$. For the Bagger-Lambert theory the supercurrent can be written as
\begin{equation}
J_\mu =  {\bar{\varepsilon}}^I J^I_\mu = \Tr (\delta \bp_A  \gamma_\mu , \psi^A ) + \Tr (\delta \bp^A \gamma_\mu , \psi_A), \label{supercurrent}
\end{equation}
where $J^{I}_\mu$ is the component supercurrent which appears in \eqref{qerd}. For future reference we write the fermion supersymmetry variations as
\begin{align}
\delta \psi_{A} &= -\G^I_{AB} \gamma^\mu D_\mu Z^B \varepsilon^I - N^I_A \varepsilon^I \nn \\
\delta \bp^A &= - \Gt^{IAB} {\bar{\varepsilon}}^I \gamma^\mu D_\mu \bz_{B} - N^{IA} \varepsilon^I \nn \\ 
\delta \psi^A &= \Gt^{IAB} \gamma^\mu D_\mu \bz_{B} \varepsilon^I - N^{IA} \varepsilon^I \\
\delta \bp_{A} &= \G^I_{AB} {\bar{\varepsilon}}^I \gamma^\mu D_\mu Z^B_A - N^I_A {\bar{\varepsilon}}^I.\nn
\end{align} 
with
\begin{align}
N^I_{A} &= \G^I_{AB} [Z^C , Z^B ; \bz_C ] - \G^I_{CD} [Z^C , Z^D ; \bz_A ] ; \\
N^{IA} &= \Gt^{IAB} [\bz_C , \bz_B ; Z^C] - \Gt^{ICD} [\bz_C , \bz_D ; Z^A].
\end{align}
We have deliberately written these variations in terms of the general 3-bracket introduced in the last section. This will result in an expression for the superalgebra in terms of 3-brackets. The benefit of this formalism is that one can easily derive the ABJM superalgebra by choosing a particular representation of the 3-bracket. The supersymmetry variation of the zeroth component of the supercurrent is computed in Appendix \ref{calc}. Since we are only interested in bosonic backgrounds we set the fermions to zero. The result is
\begin{align}
\delta J^{0,I} =  &-2\delta^{IJ} T^0_\mu \gamma^\mu \varepsilon^J + 2\delta^{IJ} V_1 \gamma^0 \varepsilon^J \nn \\
&+2 \delta^{IJ} ( \Tr (D_i \bz_B , [Z^D , Z^B ; \bz_D] ) -  \Tr (D_i Z^B , [\bz_D , \bz_B ; Z^D] ) \varepsilon^{ij}  \gamma^j \varepsilon^J \nn \\
&- \G^{C[IJ]}_{B}\Tr ( D_i Z^B , D_j \bz_C  )\varepsilon^{ij} \gamma^0 \varepsilon^J \nn \\
&- \G^{C[IJ]}_{B}( \Tr (D_0 \bz_A , [Z^B, Z^A ; \bz_C ] ) + \Tr ( D_0 Z^A, [\bz_C , \bz_A ; Z^B] )) \varepsilon^J \nn \\
&+ \G^{CD(IJ)}_{AB}( \Tr ( D^i \bz^B , [\bz_C , \bz_D ; Z^A] ) - \Tr ( D_i \bz_D ,[Z^A , Z^B ; \bz_C ] ))  \varepsilon^{ij} \gamma^j \varepsilon^J \nn \\
&- \G^{EF(IJ)}_{AB}\Tr ( [Z^C, Z^B ; \bz_C ], [\bz_E , \bz_F ; Z^A ])\varepsilon^J  \nn \\
&- \G^{EF(IJ)}_{AB}\Tr ( [Z^A , Z^B ; \bz_E ],[\bz_C , \bz_F ; Z^C])\varepsilon^J \nn  \\
&+ \G^{EF(IJ)}_{AB}\Tr ( [Z^A, Z^B ; \bz_C ],[\bz_E, \bz_F ; Z^C])\varepsilon^J , \nn
\end{align}
where we have defined
\begin{align}
\G^{CD(IJ)}_{AB} &= \G^I_{AB} \Gt^{JCD} + \Gt^{ICD} \G^J_{AB} ; \\
\Gt^{A[IJ]}_{D} &= \G^I_{DE} \Gt^{JAE} - \Gt^{IAE} \G^J_{DE}.
\end{align}
In order to determine the superalgebra from this expression we need to integrate $\delta J^{0, I}$ over the spatial worldvolume coordinates, and pull off the supersymmetry parameters $\varepsilon^J$, remembering that for Majorana spinors ${\bar{\varepsilon}} = \varepsilon^T C $. We know that $\int d^2 \sigma T^0_{\mu} = P_\mu$ so we see that the first term above will give us the usual momentum term. The other terms will form the central charges. We can write the superalgebra as
\begin{align}
\{ Q_\alpha^I , Q_\beta^J \} = &-2\delta^{IJ} (P_\mu (\gamma^\mu C)_{\alpha \beta}+ {\mathcal{Z}}_i(\gamma^i C)_{\alpha \beta}  -V_1 (\gamma^0 C)_{\alpha \beta} ) \nn \\
&-  \Gamma^{C[IJ]}_{B} ( {\mathcal{Z}}^B_{C,0} (C)_{\alpha \beta} +  {\mathcal{Z}}^{B}_{C} (\gamma^0 C )_{\alpha \beta} ) \label{minge} \\
&+ \Gamma^{EF(IJ)}_{AB}( {\mathcal{Z}}^{AB}_{EF,i} (\gamma^iC)_{\alpha \beta} \nn + {\mathcal{Z}}^{AB}_{EF} (\gamma^0 C)_{\alpha \beta}) \nn
\end{align}
where $\alpha , \beta$ are spinor indices and $i= x^1, x^2$ are the spatial coordinates of the worldvolume. The central charges are given by
\begin{align}
{\mathcal{Z}}_i &= \int d^2 \sigma \Tr (D_i \bz_B , [Z^D , Z^B ; \bz_D] ) -  \Tr (D_i Z^B , [\bz_D , \bz_B ; Z^D] )\varepsilon^{ij} \label{Z1}\\
\nn \\
{\mathcal{Z}}^B_{C} &= \int d^2 \sigma \Tr ( D_i Z^B , D_j \bz_C  )\varepsilon^{ij} \\
\nn \\
{\mathcal{Z}}^B_{C,0} &= \int d^2 \sigma ( \Tr (D_0 \bz_A , [Z^B, Z^A ; \bz_C ] ) + \Tr ( D_0 Z^A, [\bz_C , \bz_A ; Z^B] ) \\
\nn \\
{\mathcal{Z}}^{AB}_{EF,i} &= \int d^2 \sigma  \Tr ( D_i \bz^B , [\bz_E , \bz_F ; Z^A] ) - \Tr ( D_i \bz_F ,[Z^A , Z^B ; \bz_E ] ) \varepsilon^{ij} \\
\nn \\
{\mathcal{Z}}^{CE}_{BA}&= \int d^2 \sigma  \Tr ( [Z^A, Z^B ; \bz_C ],[\bz_E, \bz_F ; Z^C]) - [Z^C, Z^B ; \bz_C ], [\bz_E , \bz_F ; Z^A ]) \nn\\
\quad \quad &- \Tr ( [Z^A , Z^B ; \bz_E ],[\bz_C , \bz_F ; Z^C]) .
\end{align}
\\
These equations represent the central charges of the extended $\mathcal{N}=6$ Bagger-Lambert theory. For the specific 3-bracket realisation \eqref{bracket}, the Bagger-Lambert theory is equivalent to the ABJM theory. We will see that when we derive the ABJM central charges,  ${\mathcal{Z}}_i$,  ${\mathcal{Z}}^{AB}_{EF,i}$ can be written as surface integrals. In other words these two terms will represent topological charges in the algebra. In the next section we determine the corresponding ABJM central charges.

\section{$\mathcal{N}= 6$ ABJM Superalgebra}
In this section we will use the result of the previous section to write down the ABJM central charges. We will use the particular form of 3-bracket defined in \eqref{bracket} to map the central charge terms of the general Bagger-Lambert theory to the ABJM  theory. This will work in the same way that the Bagger-Lambert Lagrangian is mapped to the ABJM Lagrangian. The structure of the superalgebra presented in \eqref{minge} remains unchanged. Only the central charge terms are affected by the 3-bracket prescription.  Firstly we define $\Tr(X,Y) = \tr(X^\dagger Y)$ and then we write the 3-bracket as $[X,Y;Z]= XZ^\dagger Y - Y Z^\dagger X$. In order to emphasise the change from the Bagger-Lambert to ABJM picture we will relabel our fields as $Z^{A \dagger} \rightarrow X^A$ and $\bz_A \rightarrow X_A$. This matches the conventions of \cite{Band}. A simple calculation results in the following central charge terms
\begin{align}
{\mathcal{Z}}_i &= \int d^2 \sigma \tr D_i(X^B X_B X^D X_D  - X_B X^B X_D X^D ) \\
\nn \\
{\mathcal{Z}}^B_{C} &= \int d^2 \sigma \tr ( D_i X^B  D_j X_C  )\varepsilon^{ij} \\
\nn \\
{\mathcal{Z}}^B_{C,0} &= \int d^2 \sigma  \tr (D_0 X_A (X^B X_C X^A - X^A X_C X^B) - D_0 X^A(X_C X^B X_A - X_A X^B X_C) )\nn\\
\nn \\
{\mathcal{Z}}^{AB}_{EF,i} &= \int d^2 \sigma  \tr  D_i (X^B X_E X^A X_F) \varepsilon^{ij} \\
\nn \\
{\mathcal{Z}}^{AB}_{EF}&= 4\int d^2 \sigma   \tr (X^A X_C X^B X_F X^C X_E - X^B X_E X^A (X_CX^CX_F - X_F X^C X_C))
\end{align}
We see that ${\mathcal{Z}}_i$ and ${\mathcal{Z}}^{AB}_{EF,i}$ take the form of surface integrals. These terms correspond to topological terms characterizing half-BPS vacuum configurations. In \cite{IJeon2} the superalgebra of the $\mathcal{N}=8$ Bagger-Lambert theory was expressed in terms of three types of central charge; $Z_{IJ}$, $Z_{iIJKL}$ and $Z_{IJKL}$. It appears that for the $N=6$ theory the analogues of these charges are ${\mathcal{Z}}^B_{C}$, ${\mathcal{Z}}^{AB}_{EF,i}$ and ${\mathcal{Z}}^{AB}_{EF}$. We refer the reader to \cite{IJeon2} for more details on the interpretation of these central charge terms. Note that the superalgebra may be re-written in terms of trace, anti-symmetric and symmetric traceless parts. In other words we may write the superalgebra
as
\begin{equation}
\{ Q^I_\alpha , Q^J_\beta \} =  \delta^{IJ} X_{\alpha \beta} + {\tilde{Z}}^{(IJ)}_{\alpha \beta} + {\tilde{Z}}^{[IJ]}_{\alpha \beta}
\end{equation}
where $X_{\alpha \beta}$ is a singlet, ${\tilde{Z}}^{(IJ)}_{\alpha \beta}$ is symmetric traceless and ${\tilde{Z}}^{[IJ]}_{\alpha \beta}$ antisymmetric in $I,J$ respectively. Explicitly we have
\begin{align}
X_{\alpha \beta} &= -2P_\mu (\gamma^\mu C)_{\alpha \beta} - \frac{4}{3} {\mathcal{Z}}_i (\gamma^i C)_{\alpha \beta}, \nn \\
 {\tilde{Z}}^{(IJ)}_{\alpha \beta} &=  ( \G^{EF(IJ)}_{AB} {\mathcal{Z}}^{AB}_{EF,i}   - \frac{2}{3} \delta^{IJ} {\mathcal{Z}}_i ) (\gamma^i C)_{\alpha \beta} + ( \G^{EF(IJ)}_{AB}{\mathcal{Z}}^{AB}_{EF} + 2 \delta^{IJ} V_1)(\gamma^0 C)_{\alpha\beta} \nn \\
{\tilde{Z}}^{[IJ]}_{\alpha \beta} &= -  \Gamma^{C[IJ]}_{B} ( {\mathcal{Z}}^B_{C,0} C_{\alpha \beta} +  {\mathcal{Z}}^{B}_{C} (\gamma^0 C )_{\alpha \beta} ).
\end{align}
It is interesting at this stage to observe what happens when we act with $\delta_{IJ}$ on the superalgebra. In this case $\G^{C[IJ]}_B = 0$ since it is antisymmetric in $I,J$ and so ${\mathcal{Z}}^B_{C}$ and ${\mathcal{Z}}^B_{C,0} $ disappear from the algebra. Similarly ${\tilde{Z}}^{(IJ)}_{\alpha \beta}= 0$ since it is symmetric traceless. This can be confirmed by  using the fact that 
\begin{equation}
\delta_{IJ} \G^{EF(IJ)}_{AB} = \G^I_{AB} \Gt^{IEF} + \Gt^{IEF} \G^I_{AB} = -4\delta^{EF}_{AB}.
\end{equation}
Thus the only term that survives is the trace part $X_{\alpha \beta}$. We can therefore write
\begin{equation}
\delta_{IJ} \{ Q^I_\alpha, Q^J_\beta \} =  -12P_\mu (\gamma^\mu C)_{\alpha \beta}  + 8 \tr \int d^2 \sigma D^i( X_AX^AX_BX^B - X^AX_AX^BX_B) \varepsilon_{ij} (\gamma^j C)_{\alpha \beta}. \label{6}
\end{equation}
We see that the trace of the algebra contains a single central charge term, namely the one-form central charge ${\mathcal{Z}}_i$. It turns out that this charge corresponds to the energy of the BPS Fuzzy-Funnel configuration calculated in \cite{KHHL}. The ABJM BPS equations can be obtained by combining the kinetic and potential terms in the Hamiltonian and rewriting the expression as a modulus squared term plus a topological term. The squared term tells us the BPS equations and the topological term tells us the energy bound of the BPS configuration when the BPS equations are satisfied. In \cite{KHHL} the ABJM potential was written as
\begin{align}
V = &\frac{4 \pi^2}{k^2} \tr (|Z^A Z^\dagger_A Z^B - Z^B Z_A Z^A - W^A W_A Z^B + Z^B W_A W^A |^2 \nn\\
&+ |W^A W_A W^B - W^B W_A W^A - Z^A Z_A W^B + W^B Z_A Z^A|^2 ) \\
&+ \frac{16\pi^2}{k^2} \tr (|\epsilon_{AC}\epsilon^{BD} W_B Z^C W_D|^2  + |\epsilon^{AC} \epsilon_{BD}Z^B W_C Z^D|^2).\nn
\end{align}
where here $Z^A$ and $W^A$ are the upper and lower two components respectively of the 4 component complex scalar $X^A$. The first two lines correspond to D-term potential pieces whereas the last line corresponds to F-term potential pieces (from the superspace perspective). In \cite{KHHL} the potential and kinetic terms were combined in two different ways, depending on whether the F-term or D-term potential is used in conjunction with the kinetic term. This leads to two sets of BPS equations. For the case in which $W^A=0$ the scalar part of the Hamiltonian only contains D-term contributions and takes the form
\begin{align}
H = &\int dx^1 ds \tr (|\partial_s Z^A + \frac{2\pi}{k}(Z^B Z_B Z^A - Z^A Z_B Z^B)|^2) \nn \\
&+\frac{\pi}{k} \tr \partial_s (Z_A Z^A Z_B Z^B - Z^A Z_A Z^B Z_B),
\end{align}
where $x^2 = s$. As usual, the first line gives the  BPS equation
\begin{equation}
\partial_s Z^A + \frac{2\pi}{k}(Z^B Z_B Z^A - Z^A Z_B Z^B) = 0,
\end{equation}
and the second line gives the energy of the system when the BPS equation is satisfied
\begin{equation}
E = \frac{\pi}{k} \tr  \int ds dx^1  \partial_s (Z_A Z^A Z_B Z^B - Z^A Z_A Z^B Z_B).
\end{equation} 
We see that the form of this expression exactly corresponds with the central charge term appearing in \eqref{6} (when $W^A = 0$). Thus we see that the physical information corresponding to the energy bound of the fuzzy funnel configuration appears in the trace expression of the algebra, and that all the other terms vanish when the trace is taken. 

\section{Bagger-Lambert BPS equations}
In this section we would like to consider the BPS equations of the $\mathcal{N}=6$ Bagger-Lambert Theory. We begin by considering the case in which two of the complex scalars are zero and look at the BPS equation resulting from $\delta \psi = 0$ as outlined in \cite{Terashima}. Re-writing the expression for $\delta \psi_A$ in terms of 3-brackets, and assuming a vanishing gauge field, we demand that
\begin{equation}
\delta \psi_B = \gamma^\mu \partial_\mu Z^A \e_{AB} + [Z^C , Z^A ; \bz_C ] \e_{AB} + [Z^C , Z^D ; \bz_B ] \e_{CD} = 0.
\end{equation}
We will assume that $Z^3 = Z^4 = 0$ and the remaining scalar fields are functions of $x^2 = s$. We thus arrive at the following two equations
\begin{align}
\gamma^2 \partial_s Z^1 \e_{12} = [Z^2 , Z^1 ; \bz_2] \e_{12} , \\
\gamma^2 \partial_s Z^2 \e_{21} = [Z^1 , Z^2 ; \bz_1] \e_{21}.
\end{align}
Given $\gamma^2 \e_{12} = \e_{12}$ we obtain the BPS equation of the general BL theory
\begin{equation}
\partial_s Z^A = [Z^B , Z^A ; \bz_B ]. \label{general}
\end{equation}
Substituting the expression \eqref{bracket} for the 3-bracket we find  
\begin{equation}
\partial_s Z^A = \frac{2\pi}{k}(Z^B {\bz_B}^\dagger Z^A - Z^A {\bz_B}^\dagger Z^B), \label{abcd}
\end{equation}
where we have identified $\lambda = \frac{2 \pi}{k}$. This is the result of \cite{Terashima}. The general BPS equation may also be derived by considering the scalar Hamiltonian when $Z^3 = Z^4 = 0$. In this case the Bagger-Lambert potential simplifies and is proportional to $\Tr ([Z^A , Z^B ; \bz_B ], [\bz_A , \bz_B ; Z^B])$. It follows from the usual Bogomoly'ni  trick that the BPS equation is given by \eqref{general}. In \cite{KHHL} a solution to the BPS equation \eqref{abcd} was presented. The general procedure for finding a solution is to the consider the ansatz in which the complex scalar fields separate into an s-dependent and s-independent part,
\begin{equation}
Z^A = f(s) G^A , \quad f(s) = \sqrt{\frac{k}{4 \pi s}} , \label{ansatz}
\end{equation}
Looking at \eqref{abcd} we see that the $G^A$ satisfy
\begin{equation}
G^A = G^B G^\dagger_B G^A - G^A G^\dagger_B G^B.
\end{equation}
This equation is solved in \cite{Gomis3}. In \cite{KHHL} the solution is interpreted as describing a fuzzy $S^3 / Z_k$. One might ask if it is possible to find a general solution corresponding to the general BPS equation \eqref{general}. Following the same procedure one might use an ansatz similar to \eqref{ansatz}. The matrices $G^A$ would then satisfy 
\begin{equation}
G^A = [G^B , G^A ; G_B]. \label{ggg}
\end{equation}
In \cite{BL3} only one class of examples of 3-bracket were given; it would be interesting to investigate the possibility of other realisations of 3-bracket and consequently other solutions to \eqref{ggg}.  So far we have only considered the situation in which half the scalar fields are set to zero. In this case the potential takes a simple form and there is a single BPS equation. We would like to consider the BPS equations of the Bagger-Lambert theory for the case in which all scalar fields are non-zero. The scalar Hamiltonian takes the form 
\begin{align}
H &= \int dx^1 ds  \Tr (\partial_s Z^A , \partial^s \bz_A ) + \frac{2}{3} \Tr (\Upsilon^{CD}_{B} , {\bar{\Upsilon}}^{B  }_{CD} ). 
\end{align}
We can write this as a sum of squares,
\begin{align}
 H &=  \int dx^1 ds \Tr | \partial_s Z^A - \frac{1}{\sqrt{3}} \varepsilon^{AB}_{\phantom{1} \phantom{1} \phantom{1}CD}\Upsilon^{CD}_{B} |^2  + \Tr |[Z^C, Z^B; \bz_C]|^2 + T_1 \label{upon}.
\end{align}
This leads to the following set of BPS equations
\begin{align}
\partial_s Z^A - \frac{1}{\sqrt{3}} \varepsilon^{AB}_{\phantom{1} \phantom{1} \phantom{1}CD}\Upsilon^{CD}_{B} &= 0 \label{xcx}\\
[Z^C, Z^B; \bz_C] &= 0.
\end{align}
Writing out \eqref{xcx} explicitly in terms of the component scalars we find expressions of the form
\begin{align}
\partial_s Z^1 &= \frac{1}{\sqrt{3}}[Z^2 , Z^3 , \bz_4 ] = \frac{1}{\sqrt{3}}[Z^4 , Z^2 , \bz_3 ] = \frac{1}{\sqrt{3}}[Z^4 , Z^3 ; \bz_2 ] \nn \\
\partial_s Z^2 &= \frac{1}{\sqrt{3}}[Z^3, Z^4 ; \bz_1 ] = \frac{1}{\sqrt{3}}[Z^1 , Z^3 , \bz_4 ] = \frac{1}{\sqrt{3}}[Z^1 , Z^4 ; \bz_3 ] \nn \\
\partial_s Z^3 &= \frac{1}{\sqrt{3}}[Z^4, Z^1 ; \bz_2 ] = \frac{1}{\sqrt{3}}[Z^2 , Z^4 , \bz_1 ] = \frac{1}{\sqrt{3}}[Z^2 , Z^1 ; \bz_4 ] \nn \\
\partial_s Z^4 &= \frac{1}{\sqrt{3}}[Z^1, Z^2 ; \bz_3 ] = \frac{1}{\sqrt{3}}[Z^3 , Z^1 , \bz_2 ] = \frac{1}{\sqrt{3}}[Z^3 , Z^2 ; \bz_2 ] .
\end{align}
Note that if we choose to set half the scalar fields to zero then any term involving the epsilon tensor will vanish and we are left with a trivial set of constraints, namely $\partial_s Z^A = [Z^C, Z^B ; Z_C] = 0$. Alternatively we can re-write \eqref{upon} as
\begin{align}
 H &=  \int dx^1 ds \Tr |\partial_s Z^A - [Z^B, Z^A; \bz_B ]|^2  + \frac{1}{3}\Tr|\varepsilon^{AB}_{\phantom{1} \phantom{1} \phantom{1}CD} \Upsilon^{CD}_B|^2 + T_2 \label{ughj}
\end{align}
which leads to the following BPS equations
\begin{align}
\partial_s Z^A - [Z^B, Z^A; \bz_B ] &= 0 \label{mum}\\
\varepsilon^{AB}_{\phantom{1} \phantom{1} \phantom{1}CD}\Upsilon^{CD}_{B} &= 0. \label{acg}
\end{align}
For the case in which half the scalars are set to zero we see that \eqref{acg} vanishes and that \eqref{mum} exactly corresponds to the BPS equation derived by setting $\delta \psi =0$. It is worth mentioning that we could have written the Hamiltonian as
\begin{align}
H &= \frac{2}{3} \int dx^1 ds |\frac{1}{2}\varepsilon^{CD}_{\phantom{1} \phantom{1} \phantom{1}BA} \partial_s Z^A - \Upsilon^{CD}_{B}|^2 + T_3, \label{BPS}
\end{align}
in which case we would have a single set of BPS equations of the form
\begin{equation}
\varepsilon^{CD}_{\phantom{1} \phantom{1} \phantom{1}BA} \partial_s Z^A - 2\Upsilon^{CD}_{B} = 0.
\end{equation}
However it is not clear how to extract \eqref{general} for the case in which half the scalars are zero. 

\section{Conclusion and Discussion}
In this paper we calculated the extended worldvolume superalgebra of the $N=6$ Bagger-Lambert Theory. With a particular choice of 3-bracket we were able to derive the ABJM superalgebra. We found that the central charge corresponding to the half-BPS fuzzy funnel configuration of the ABJM theory appears as a diagonal element of the superalgebra. It would be interesting to study the off-diagonal central charge terms and provide a physical interpretation. It may be possible to re-write the superlgebra in a neater form by using the equations of motion (as was done for the $\mathcal{N} = 8$ in \cite{Pass}). This may simplify the structure of the central charge terms allowing for easier interpretation. It is interesting to note that ${\mathcal{Z}}_i$ exactly corresponds with the topological term appearing in \cite{KHHL} when the kinetic term is combined with the F-term potential piece. Furthermore it appears that ${\mathcal{Z}}^{AB}_{EF,i}$ has the same structure as the topological term corresponding to the D-term configuration. Thus it would seem that these two central charge terms characterise the topological information corresponding to the two sets of BPS equations appearing in \cite{KHHL}.
\\
\\
In this paper we have also derived two sets of BPS equations for the general $\mathcal{N}=6$ Bagger-Lambert theory. For the case in which half the scalars are set to zero we recover the half-BPS result derived by setting $\delta \psi =0$. It would be interesting to try and find solutions to these equations in the case where more than half the scalar fields are active. Related to this is the question of whether its possible to write the Bagger-Lambert scalar Hamiltonian as 
\begin{align}
 H &=  \int dx^1 ds \Tr ( \partial_s Z^A - g^{AB}_{\phantom{1} \phantom{1} \phantom{1}CD}\Upsilon^{CD}_{B} )^2  + T \label{upo}
\end{align}
with the condition that
\begin{equation}
g^{AB}_{\phantom{1} \phantom{1} \phantom{1}CD}g_{AE}^{\phantom{1} \phantom{1} \phantom{1}FG} \Tr (\Upsilon^{CD}_{B} , {\bar{\Upsilon}}^{E}_{FG} ) = \frac{2}{3}\Tr (\Upsilon^{CD}_{B} , {\bar{\Upsilon}}^{B}_{CD} ) \label{const}
\end{equation}
where T is a topological term. If this constraint is satisfied then we have a set of BPS equations of the form
\begin{equation}
\partial_s Z^A - \kappa g^{AB}_{\phantom{1} \phantom{1} \phantom{1}CD}\Upsilon^{CD}_{B} = 0 \label{BPS1}
\end{equation}
where $A,B = 1, \ldots 4$. It is interesting to note that the constraint \eqref{const} is analogous to the situation encountered when considering M5-brane calibrations \cite{Berman, Krishnan}. In the case of the $\mathcal{N}=8$ Bagger-Lambert theory the constraint takes the form 
\begin{equation}
\frac{1}{3!} g_{IJKL} g_{IPQR} \Tr ([X^J, X^K, X^L],[X^P, X^Q, X^R]) = \Tr([X^I, X^J, X^K],[X^I, X^J, X^K]).
\end{equation}
The $g_{IJKL}$ are related to the calibrating forms of the cycle on which the M5-brane wraps and are therefore completely antisymmetric in their indices. For the case in which only half the scalar fields are activated it is possible to solve the constraint by writing $ g_{IJKL} = \varepsilon_{IJKL} $. This choice corresponds to a fuzzy-funnel configuration in which multiple M2-branes expand into a single M5-brane, and is described by the standard Basu-Harvey equation. For the situation in which more scalars are activated, additional constraints arise which have to be imposed alongside the Basu-Harvey equation. It would be interesting to see whether the results of \cite{Berman} can be derived from the ABJM theory. We leave this for future work.

\section*{Acknowledgements\markboth{Acknowledgements}{Acknowledgements}} 
We would like to thank B. Spence for reading a draft version of this paper and for useful discussions during its completion. We are also grateful to D. Thompson, D. Berman and V. Calo for helpful comments and clarifications. AL is supported by an STFC grant.

\begin{appendix}
\appendix
\section{Conventions and Useful Information} \label{conventions}

In what follows we will need to make use of the following information. The supersymmetry parameters of the $N=6$ ABJM theory transform in the $6$ representation of $SU(4)$. We can write the susy parameter $\varepsilon_{AB}$ in terms of a basis of  $4 \times 4$ gamma matrices as
\begin{equation}
\varepsilon_{AB} = \epsilon^I .(\G^I_{AB}),
\end{equation} 
with $I= 1, \ldots 6$. The gamma matrices are antisymmetric $(\G^I_{AB} = -\G^I_{BA})$ and satisfy the following relation
\begin{equation}
\G^I_{AB} \Gt^{JBC} + \G^J_{AB} \Gt^{IBC} = 2 \delta^{IJ} \delta^C_A \label{bew}
\end{equation}
where
\begin{equation}
\Gt^{IAB} = \frac{1}{2} \varepsilon^{ABCD} \G^I_{CD} = - (\G^I_{AB})^*. \label{up}
\end{equation}
We note that the $4 \times 4$ matrices $\G^I$ act on a different vector space to the $2 \times 2$ matrices $\gamma^\mu$ which are defined as world volume gamma matrices. These two types of gamma matrix commute with one another. It is also important to note the following relations
\begin{align}
\G^I_{AB} \Gt^{ICD} &= -2 \delta^{CD}_{AB} = -2 (\delta^C_A \delta^D_B - \delta^C_B \delta^D_A ) \label{qed}\\ 
\G^I_{AB} \Gt^{IBD} &= 6 \delta^D_A. \label{qed2}
\end{align}
Acting with $\varepsilon^{ABMN} \varepsilon_{CDPQ}$ on both sides of \eqref{qed} one can show that
\begin{equation}
\Gt^{ICD}\G^I_{AB} = - 2 \delta^{CD}_{AB}.
\end{equation}
It therefore follows that
\begin{align}
\G^I_{AB} \Gt^{ICD} + \Gt^{ICD} \G^I_{AB} &= -4\delta^{CD}_{AB} \label{nonzero}\\
\G^I_{AB} \Gt^{ICD} - \Gt^{ICD} \G^I_{AB} &= 0 \label{zero}
\end{align}
We will also need the following identity in what follows
\begin{align}
\G^I_{AB} \Gt^{AC} + \Gt^{IAC} \G^J_{AB}  &= \G^I_{AB} \Gt^{AC} + \frac{1}{4} \varepsilon^{ACDE} \varepsilon_{ABFG} \G^I_{DE} \Gt^{JFG} \nn \\
&= \frac{1}{2} \delta^C_B \G^I_{FG} \Gt^{JFG} = 2 \delta^{IJ} \delta^C_B. \label{apefarm}
\end{align}
and therefore
\begin{equation}
\G^I_{FG} \Gt^{JFG} = 4 \delta^{IJ}.
\end{equation}
Note that in obtaining the last line of \eqref{apefarm} we made use of \eqref{bew} and the epsilon tensor identity
\begin{align}
\varepsilon^{ACDE} \varepsilon_{ABFG} = &+ \delta^C_B \delta^D_F \delta^E_G + \delta^C_F \delta^D_G \delta^E_B + \delta^C_G \delta^D_B \delta^E_F \nn \\
&- \delta^C_B \delta^D_G \delta^E_F - \delta^C_F \delta^D_B \delta^E_G - \delta^C_G \delta^D_F \delta^E_B  
\end{align}
Similarly we have
\begin{equation}
\G^I_{AB} \Gt^{AC} - \Gt^{IAC} \G^J_{AB} = 2 \G^I_{AB} \Gt^{AC} - \frac{1}{2} \delta^C_B \G^I_{FG} \Gt^{JFG}.\label{1d2}
\end{equation}
It is possible to derive identities involving $\e_{AB}$ based on the relations between the basis gamma matrices $\Gamma^I$. In \cite{BL3} Bagger and Lambert make use of the following identities
\begin{equation}
\frac{1}{2}\be^{CD}_1 \gamma_\nu \e_{2CD} \delta^A_B = \be^{AC}_1 \gamma_\nu \e_{2BC} - \be^{AC}_2 \gamma_\nu \e_{1BC} \label{iden}
\end{equation}
and
\begin{align}
2 \be^{AC}_1 \e_{2BD} - 2 \be^{AC}_2 \e_{1BD} = &+ \be^{CE}_1 \e_{2DE} \delta^A_B - \be^{CE}_2 \e_{1DE} \delta^A_B \nn \\
&- \be^{AE}_1 \e_{2DE} \delta^C_B + \be^{AE}_2 \e_{1DE} \delta^C_B \nn \\
&+ \be^{AE}_1 \e_{2BE} \delta^C_D - \be^{AE}_2 \e_{1BE} \delta^C_D  \label{swiss} \\
&- \be^{CE}_1 \e_{2BE} \delta^A_D + \be^{CE}_2 \e_{1BE} \delta^A_D. \nn
\end{align}
Both of these identities can be re-written in terms of identities involving the Majorana spinors $\epsilon^I$ and the gamma matrices $\Gamma^I$.

\section{Determination Of Surface Term}\label{surface}
In this appendix we show explicitly how to calculate the surface term $V^\mu $ associated with the Lagrangian \eqref{zop}. Only certain parts of the variation of the Lagrangian contribute to the surface terms, namely those kinetic and coupling terms which upon variation contain derivatives. Lets look at each part of the Lagrangian in turn.

\subsection{Kinetic Term}
\begin{equation}
\mathcal{L}_{kinetic} = -D^\mu \bz^a_A D_\mu Z^A_a - i \bp^{Aa} \gamma^\mu D_\mu \psi_{Aa}.
\end{equation}
Varying the kinetic terms one has
\begin{align}
\delta \mathcal{L}_{kinetic}=&- \overbrace{h^{\bb a} D^\mu (\delta \bz_{A \bb}) D_\mu Z^A_a}^1 + \overbrace{h^{\bb a} \delta {\tilde{A}}^{* \bc}_{\mu \phantom{1} \bd} \bz_{A \bc} D_\mu Z^A_a}^2 \nonumber \\
 &- \overbrace{D^\mu \bz^a_A D_\mu (\delta Z^A_a)}^3 + \overbrace{ h^{d \bd} D^\mu \bz_{A\bd} \delta {\tilde{A}}^{\phantom{1}c}_{\mu \phantom{1} d} Z^A_c }^4\\
&- \overbrace{i \delta \bp^{Aa} \gamma^\mu D_\mu \psi_{Aa}}^5 - \overbrace{i \bp^{Aa} \gamma^\mu D_\mu (\delta \psi_{Aa})}^6 + \overbrace{i \bp^{Aa} \gamma^\mu \delta{\tilde{A}}^{\phantom{1}b}_{\mu \phantom{1} a} \psi_{Ab}}^7 \nonumber
\end{align}
Inserting the supersymmetry transformations into the above one finds the following terms
\begin{align}
1) &=-i D^\mu \bp^{Ba} D_\mu Z^A_a \e_{AB} \nonumber \\
2) &=+i \be^{AB} \gamma_\mu D^\mu Z^C_b \bz_{B \bd} \bz_{C \bc} \psi_{A a} f^{ab \bc \bd} - i \bp^B_{\ba} \gamma_\mu Z^A_b \bz_{C \bc} D_\mu Z^C_a f^{ba \bc \ba} \e_{AB} \nonumber \\
3) &=- i \be^{AB} D^\mu \bz^a_A D_\mu \psi_{Ba} \nonumber \\
4) &=-i \be^{AB} \gamma^\mu D_\mu \bz_{C \bc} \bz_{B \bd} Z^C_b \psi_{Aa} f^{ab \bc \bd} + i \bp^B_{\bb} \gamma_\mu D^\mu \bz_{C \bd} Z^A_a Z^C_c f^{ca \bb \bd} \e_{AB} \\
5) &=- i \be^{AB} D_\mu \bz^a_B \gamma^\mu \gamma^\nu D_\nu \psi_{Aa} - i\be^{AB} \bz_{C \bc} \bz_{B \bd} Z^C_b \gamma^\mu D_\mu \psi_{Aa} f^{ab \bc \bd}\nonumber \\
&\quad - i \be^{CD} \gamma^\mu D_\mu \psi_{Aa} Z^A_c \bz_{D \bb} \bz_{C \bd} f^{ca \bd \bb} \nonumber \\
6) &= -i\bp^{A}_{\ba} \gamma^\mu \gamma^\nu D_\mu D_\nu Z^B_a \e_{BA} - i \bp^{Aa} \gamma^\mu D_\mu (Z^C_d Z^B_b \bz_{C \bc})f^{db\bc \ba} \e_{BA}\nonumber\\
&\quad - i \bp^{A}_{\ba} \gamma^\mu D_\mu (Z^C_d Z^D_b \bz_{A \bc}) f^{db \bc \ba} \e_{CD}.\nonumber
\end{align}
We don't include $7)$ above as this term contains no derivatives and therefore won't contribute to the surface terms.
\subsection{Coupling terms}
\begin{equation}
\mathcal{L}_{coupling} = \mathcal{L}_{(1)} + \mathcal{L}_{(2)}
\end{equation}
where
\begin{align}
\mathcal{L}_{(1)} &= -i f^{ab \bc \bd} \bp^A_{\bd} \psi_{Aa} Z^B_b \bz_{B \bc} + 2i f^{ab \bc \bd} \bp^A_{\bd} \psi_{Ba} Z^B_b \bz_{A \bc}.\\
\mathcal{L}_{(2)} &= \frac{i}{2} \varepsilon_{ABCD} f^{ab \bc \bd} \bp^A_{\bd} \psi^B_{\bc} Z^C_a Z^D_b - \frac{i}{2} \varepsilon^{ABCD} f^{cd \ba \bb} \bp_{Ac} \psi_{Bd} \bz_{C \ba} \bz_{D \bb}.
\end{align}
We will tackle each in turn

\begin{align}
\delta\mathcal{L}_{(1)} = &-i \overbrace{f^{ab \bc \bd} \delta \bp^A_{\bd} \psi_{Aa} Z^B_b \bz_{B \bc}}^1 -i \overbrace{f^{ab \bc \bd} \bp^A_{\bd} \delta \psi_{Aa} Z^B_b \bz_{B \bc}}^2 \nonumber \\
&+ 2i \overbrace{f^{ab \bc \bd} \delta\bp^A_{\bd} \psi_{Ba} Z^B_b \bz_{A \bc}}^3 + 2i \overbrace{f^{ab \bc \bd} \bp^A_{\bd} \delta \psi_{Ba} Z^B_b \bz_{A \bc}}^4
\end{align}
Inserting the supersymmetry transformations into this we have
\begin{align}
1) &= - i \be^{AB} \gamma^\mu D_\mu \bz_{B \bd} \bz_{C \bc} Z^C_b \psi_{Aa} f^{ab \bc \bd} \nonumber\\
2) &= - i \bp^{A}_{\bd} \gamma^\mu D_\mu Z^C_a Z^B_b \bz_{B \bc} f^{ab \bc \bd} \e_{CA} \nonumber\\
3) &= 2i \be^{AB} \gamma^\mu D_\mu \bz_{B \bd} \bz_{A \bc} Z^C_b \psi_{Ca} f^{ab \bc \bd} \\
4) &= 2i \bp^A_{\bd} \gamma^\mu D_\mu Z^C_a Z^B_b \bz_{A \bc} f^{ab \bc \bd} \e_{CB} \nonumber 
\end{align}
For $\delta \mathcal{L}_{(2)}$ we find
\begin{align}
\delta \mathcal{L}_{(2)} =& \overbrace{\frac{i}{2} \varepsilon_{ABCD} f^{ab \bc \bd} \delta \bp^A_{\bd} \psi^B_{\bc} Z^C_a Z^D_b}^5 + \overbrace{\frac{i}{2} \varepsilon_{ABCD} f^{ab \bc \bd} \bp^A_{\bd} \delta\psi^B_{\bc} Z^C_a Z^D_b}^6 \nonumber\\
&- \overbrace{\frac{i}{2} \varepsilon^{ABCD} f^{cd \ba \bb} \delta \bp_{Ac} \psi_{Bd} \bz_{C \ba} \bz_{D \bb}}^7 - \overbrace{\frac{i}{2} \varepsilon^{ABCD} f^{cd \ba \bb} \bp_{Ac} \delta \psi_{Bd} \bz_{C \ba} \bz_{D \bb}}^8 
\end{align}
Inserting the supersymmetry variations one finds
\begin{align}
5) &= \frac{i}{2} \be_{CD} \gamma^\mu D_\mu \bz_{B \bd} Z_a^C Z^D_b \psi^B_{\bc} f^{a b \bc \bd} + i \be_{BC} \gamma^\mu D_\mu \bz_{D \bd} Z^C_a Z^D_b \psi^B_{\bc} f^{ab \bc \bd} \nonumber\\
6) &= \frac{i}{2} \bp^A_{\bd} \gamma^\mu D_\mu \bz_{A \bc} Z^C_a Z^D_b f^{ab \bc \bd} \e_{CD} + i \bp^A_{\bd} \gamma^\mu D_\mu \bz_{D \bc} Z^C_a Z^D_b f^{ab \bc \bd} \e_{AC} \nonumber \\
7) &= \frac{i}{2} \be^{CD} \gamma^\mu D_\mu Z^B_c \bz_{C \ba} \bz_{D \bb} \psi_{B d} f^{c d \ba \bb} + i \be^{BC} \gamma^\mu D_\mu Z^D_c \bz_{C \ba} \bz_{D \bb} \psi_{B d} f^{cd \ba \bb} \\
8) &= \frac{i}{2} \bp_{Ac} \gamma^\mu D_\mu Z^A \bz_{C \ba} \bz_{D \bb} f^{cd \ba \bb}\e^{CD} + i \bp_{Ac} \gamma^\mu D_\mu Z^D_b \bz_{C \ba} \bz_{D \bb} f^{cd \ba \bb} \e^{AC} , \nonumber 
\end{align}
where in determining the above expressions we made use of the reality condition $\e^{AB} = \frac{1}{2} \varepsilon^{ABCD} \e_{CD}$ .We also found the following epsilon tensor identity useful
\begin{align}
\varepsilon^{ABCD} \varepsilon_{AEFG} = &+\delta^B_E \delta^C_F \delta^D_G + \delta^B_F \delta^C_G \delta^D_E + \delta^B_G \delta^C_E \delta^B_F  \nonumber \\
&-\delta^B_E \delta^C_G \delta^D_F - \delta^B_G \delta^C_F \delta^D_E - \delta^B_F \delta^C_E \delta^B_G.
\end{align}
\subsection{Terms in $\be^{AB}$}
We now gather all those terms of the form $\be^{AB}$,
\\
\begin{align}
\delta {\mathcal{L}}_{\be} = &+ i \be^{AB} \gamma_\mu D^\mu Z^C_b \bz_{B \bd} \bz_{C \bc} \psi_{A a} f^{ab \bc \bd} - i \be^{AB} D^\mu \bz^a_A D_\mu \psi_{Ba} \nonumber \\
&- i \be^{AB} \gamma^\mu D_\mu \bz_{C \bc} \bz_{B \bd} Z^C_b \psi_{Aa} f^{ab \bc \bd} - i \be^{AB} D_\mu \bz^a_B \gamma^\mu \gamma^\nu D_\nu \psi_{Aa} \nonumber \\
 &-  i\be^{AB} \bz_{C \bc} \bz_{B \bd} Z^C_b \gamma^\mu D_\mu \psi_{Aa} f^{ab \bc \bd} - i \be^{CD} \gamma^\mu D_\mu \psi_{Aa} Z^A_c \bz_{D \bb} \bz_{C \bd} f^{ca \bd \bb} \\
&-  i \be^{AB} \gamma^\mu D_\mu \bz_{B \bd} \bz_{C \bc} Z^C_b \psi_{Aa} f^{ab \bc \bd} +2i \be^{AB} \gamma^\mu D_\mu \bz_{B \bd} \bz_{A \bc} Z^C_b \psi_{Ca} f^{ab \bc \bd} \nonumber \\
&+  i \be^{CD} \gamma^\mu D_\mu Z^B_c \bz_{C \ba} \bz_{D \bb} \psi_{B d} f^{c d \ba \bb} + 2i \be^{BC} \gamma^\mu D_\mu Z^D_c \bz_{C \ba} \bz_{D \bb} \psi_{B d} f^{cd \ba \bb} \nonumber
\end{align}
\\
where in the last line we have combined terms in $7)$ and $8)$ by making use of the fact that $\be^{CD} \gamma^\mu \psi_{Bd} = - \bp_{Bd} \gamma^\mu \e^{CD}$. All the terms of order $Z \bz \bz$ combine into two total derivatives. Thus we are left with
\\
\begin{align}
\delta {\mathcal{L}}_{\be} = &+ D_\mu (i \be^{AB} \gamma^\mu Z^C_b \bz_{A \bc} \bz_{B \bd} \psi _{Ca} f^{ab \bc \bd} - i\be^{AB} \gamma^\mu Z^C_b \bz_{B \bd} \bz_{C \bc} \psi_{Aa} f^{ab \bc \bd}) \nonumber \\
&-  i \be^{AB} D^\mu \bz^a_A D_\mu \psi_{Ba} - i \be^{AB} D_\mu \bz^a_B \gamma^\mu \gamma^\nu D_\nu \psi_{Aa}
\end{align}
\\
We can write these last two terms as a total derivative, plus a piece proportional to the gauge field strength. Thus we finally arrive at
\\
\begin{align}
\delta {\mathcal{L}}_{\be} =  & D_\mu (i \be^{AB} \gamma^\mu Z^C_b \bz_{A \bc} \bz_{B \bd} \psi _{Ca} f^{ab \bc \bd} - i\be^{AB} \gamma^\mu Z^C_b \bz_{B \bd} \bz_{C \bc} \psi_{Aa} f^{ab \bc \bd} \nonumber\\
&- i \be^{AB} D^\mu \bz^a_A \psi_{Ba} - i \be^{AB} D_\nu \bz^a_B \gamma^\nu \gamma^\mu \psi_{Aa} )
\end{align}
\subsection{Terms in $\e_{AB}$}
Gathering all the terms of the form $\e_{AB}$ we find
\begin{align}
 \delta {\mathcal{L}}_{\e} = &-i D^\mu \bp^{Ba} D_\mu Z^A_a \e_{AB}-i\bp^{A}_{\ba} \gamma^\mu \gamma^\nu D_\mu D_\nu Z^B_a \e_{BA} \nonumber\\
 &+ i \bp^B_{\bb} \gamma_\mu D^\mu \bz_{C \bd} Z^A_a Z^C_c f^{ca \bb \bd} \e_{AB} - i \bp^B_{\ba} \gamma_\mu Z^A_b \bz_{C \bc} D_\mu Z^C_a f^{ba \bc \ba} \e_{AB}\nonumber \\
& - i \bp^{Aa} \gamma^\mu D_\mu (Z^C_d Z^B_b \bz_{C \bc})f^{db\bc \ba} - i \bp^{A}_{\ba} \gamma^\mu D_\mu (Z^C_d Z^D_b \bz_{A \bc}) f^{db \bc \ba} \e_{CD}\\
&- i \bp^{A}_{\bd} \gamma^\mu D_\mu Z^C_a Z^B_b \bz_{B \bc} f^{ab \bc \bd} \e_{CA} + 2i \bp^A_{\bd} \gamma^\mu D_\mu Z^C_a Z^B_b \bz_{A \bc} f^{ab \bc \bd} \e_{CB} \nonumber \\
& +i \bp^A_{\bd} \gamma^\mu D_\mu \bz_{A \bc} Z^C_a Z^D_b f^{ab \bc \bd} \e_{CD} + 2i \bp^A_{\bd} \gamma^\mu D_\mu \bz_{D \bc} Z^C_a Z^D_b f^{ab \bc \bd} \e_{AC} \nonumber
\end{align}
A simple re-labeling of the indices reveals that all the terms containing $ZZ\bz$ vanish identically leaving
\begin{equation}
 \delta {\mathcal{L}}_{\e} = -i D^\mu \bp^{Ba} D_\mu Z^A_a \e_{AB}+ i\bp^{A}_{\ba} \gamma^\mu \gamma^\nu D_\mu D_\nu Z^B_a \e_{AB}
\end{equation}
and we can re-write this as a total derivative
\begin{equation}
 \delta {\mathcal{L}}_{\e} = D_\mu (-i \bp^{Ba} D_\mu Z^A_a \e_{AB})
\end{equation}
Combining the results of the previous two sub-sections we find,
\begin{equation}
\delta \mathcal{L} = \delta {\mathcal{L}}_{\be} + \delta {\mathcal{L}}_{\e} = \partial_\mu V^\mu
\end{equation}
with $V^\mu$ given by
\begin{align}
 V^\mu = &- i \be^{AB} D^\mu \bz^a_A \psi_{Ba} - i\bp^{Ba} D^\mu Z^A_a \e_{AB} - i \be^{AB} D_\nu \bz^a_B \gamma^\nu \gamma^\mu \psi_{Aa} \nonumber \\
&- i \be^{AB} \gamma^\mu Z^C_b \bz_{B \bd} \bz_{C \bc} \psi_{Aa} f^{ab\bc \bd} + i \be^{AB} \gamma^\mu Z^C_b \bz_{A\bc} \bz_{B \bd} \psi_{Ca} f^{ab\bc \bd}.
\end{align}
\section{Bagger-Lambert Superalgebra Calculations} \label{calc}
In this section we calculate the supersymmetric variation of $J^{0, I}$. Given the supercurrent expression \eqref{supercurrent} one finds
\begin{align}
\delta J^{0,I} = &+ \overbrace{\Tr (\G^I_{AB} \Gt^{JAC}\gamma^\nu \gamma^0 \gamma^\rho  D_\nu Z^B  , D_\rho \bz_C \varepsilon^J ) + \Tr (\Gt^{IAB} \G^J_{AC} \gamma^\nu \gamma^0 \gamma^\rho D_\nu \bz_B , D_\rho Z^C)}^{(a)} \nn \\
&- \overbrace{\Tr (\G^I_{AB} \gamma^\nu \gamma^0 D_\nu Z^B , N^{JA} \varepsilon^J ) - \Tr (N^I_A \Gt^{JAC} \gamma^0 \gamma^\rho,  D_\rho \bz_C \varepsilon^J )}^{(b)} \nn \\
&+ \Tr (\Gt^{IAB} \gamma^\nu \gamma^0 D_\nu \bz_B, N^J_A \varepsilon^J ) + \Tr(N^{IA} \Gamma^J_{AC} \gamma^0 \gamma^\rho D_\rho Z^C \varepsilon^J ) \\
&+ \overbrace{\Tr (N^I_A \gamma^0 , N^{JA} \varepsilon^J ) + \Tr (N^{IA} \gamma_0 , N^J_A \varepsilon^J)}^{(c)}. \nn
\end{align}
\subsection{(a) terms}
The $(a)$ terms may be written as
\begin{align}
(a) = &-\Tr ((\G^I_{AB} \Gt^{JAC} + \Gt^{IAC} \G^J_{AB})\gamma^0 D_0 Z^B , D_0 \bz_C \varepsilon^J ) \nn \\
&-\Tr ((\G^I_{AB} \Gt^{JAC} + \Gt^{IAC} \G^J_{AB})\gamma^i D_0 Z^B , D_i \bz_C \varepsilon^J ) \nn \\
&-\Tr ((\G^I_{AB} \Gt^{JAC} + \Gt^{IAC} \G^J_{AB})\gamma^i D_i Z^B , D_0 \bz_C \varepsilon^J ) \nn \\
&- \Tr ((\G^I_{AB} \Gt^{JAC} + \Gt^{IAC} \G^J_{AB}) \gamma^0 D_i Z^B , D^i \bz_C \varepsilon^J ) \\
&- \Tr ((\G^I_{AB} \Gt^{JAC} - \Gt^{IAC} \G^J_{AB})\gamma^{ij} \gamma^0 D_i Z^B , D_j \bz_C \varepsilon^J ) \nn .
\end{align}
The first four terms can be further simplified by using the relation \eqref{apefarm}.
\begin{align}
(a) = &-2\delta^{IJ} \Tr(\gamma^0 D_0 Z^B , D^0 \bz_B \varepsilon^J) -2 \delta^{IJ} \Tr( \gamma^0 D_i Z^B , D^i \bz_B \varepsilon^J ) \nn \\
&-2 \delta^{IJ} \Tr (\gamma^i D_0 Z^B , D_i \bz_B \varepsilon^J ) -2 \delta^{IJ} \Tr( \gamma^i D_i Z^B , D_0 \bz_B \varepsilon^J ) \\
&- \Tr ((\G^I_{AB} \Gt^{JAC} - \Gt^{IAC} \G^J_{AB})\gamma^{ij} \gamma^0 D_i Z^B , D_j \bz_C \varepsilon^J ) \nn.
\end{align}
\subsection{(b) terms}
The $(b)$ terms may be written as
\begin{align}
(b) = &-2 \delta^{IJ} \Tr (D^i \bz_B , [Z^D , Z^B ; \bz_D] \gamma^0 \gamma^i \varepsilon^J ) + 2 \delta^{IJ} \Tr (D^i Z^B , [\bz_D , \bz_B ; Z^D] \gamma^0 \gamma^i \varepsilon^J ) \nn \\
&+ \Tr((\G^I_{DE} \Gt^{JAC} + \Gt^{IAC} \G^J_{DE}) [Z^D , Z^E ; \bz_A ] \gamma^0 \gamma^i , D_i \bz_C \varepsilon^J ) \nn \\
&- \Tr ((\G^I_{AB} \Gt^{JCD} + \Gt^{ICD} \G^J_{AB})[\bz_C , \bz_D ; Z^A] \gamma^0 \gamma^i ,  D_i Z^B \varepsilon^J )\nn  \\
&+ \Tr ((\G^I_{AB} \Gt^{JAC} - \Gt^{IAC} \G^J_{AB}) [Z^D , Z^B ; \bz_D ] , D_0 \bz_C \varepsilon^J ) \nn \\
&+ \Tr ((\G^I_{AB} \Gt^{JAC} - \Gt^{IAC} \G^J_{AB}) [\bz_D , \bz_C ; Z^D] , D_0 Z^B \varepsilon^J ) \\
&- \Tr ((\G^I_{AB} \Gt^{JAC} - \Gt^{IAC} \G^J_{AB}) [Z^D , Z^E ; \bz_A ] , D_0 \bz_C \varepsilon^J )\nn \\
&- \Tr ((\G^I_{AB} \Gt^{JAC} - \Gt^{IAC} \G^J_{AB}) [\bz_C , \bz_D ; Z^A] , D_0 Z^B \varepsilon^J )\nn 
\end{align}
The terms involving $D_0$ can be greatly simplified by using \eqref{swiss}. After a bit of rearrangement and relabeling we can write the (b) terms as
\begin{align}
(b) = &+2 \delta^{IJ} \Tr (D^i \bz_B , [Z^D , Z^B ; \bz_D] \varepsilon^{ij} \varepsilon^J ) - 2 \delta^{IJ} \Tr (D^i Z^B , [\bz_D , \bz_B ; Z^D] \varepsilon^{ij} \varepsilon^J ) \nn \\
&+ \Tr (\G^{CD(IJ)}_{AB} D^i \bz^B , [\bz_C , \bz_D ; Z^A] \varepsilon^{ij} \gamma^j \varepsilon^J ) - \Tr (\G^{CD(IJ)}_{AB} D_i \bz_D [Z^A , Z^B ; \bz_C ] \varepsilon^{ij} \gamma^j \varepsilon^J ) \nn \\
&- \Tr (\Gt^{AE[IJ]}_{DE} D_0 \bz_C , [Z^D, Z^C ; \bz_A ] \varepsilon^J ) - \Tr (\Gt^{AE[IJ]}_{DE} D_0 Z^C [\bz_A , \bz_C ; Z^D] \varepsilon^J ),
\end{align}
where
\begin{align}
\G^{CD(IJ)}_{AB} &= \G^I_{AB} \Gt^{JCD} + \Gt^{ICD} \G^J_{AB} ; \\
\Gt^{A[IJ]}_{D} &= \G^I_{DE} \Gt^{JAE} - \Gt^{IAE} \G^J_{DE},
\end{align}
and we have used the fact that in 3 dimensions $\gamma^{ij} \propto \varepsilon^{ij}$. We have also used the fact that $\gamma^0 \gamma^i = - \varepsilon^{ij} \gamma^{012}$ and $\gamma^{012} \varepsilon^J = \varepsilon^J$.
\subsection{(c) terms}
The (c) terms may be written as
\begin{align}
(c) = &-2 \delta^{IJ} \Tr ([Z^C , Z^B ; \bz_C ],[\bz_F , \bz_B ; Z^F ])\varepsilon^J \nn \\
&- \G^{EF(IJ)}_{AB}\Tr ( [Z^C, Z^B ; \bz_C ], [\bz_E , \bz_F ; Z^A ])\varepsilon^J \nn \\
&- \G^{EF(IJ)}_{AB}\Tr ( [Z^A , Z^B ; \bz_E ],[\bz_C , \bz_F ; Z^C])\varepsilon^J \\
&+ \G^{EF(IJ)}_{AB}\Tr ( [Z^A, Z^B ; \bz_C ],[\bz_E, \bz_F ; Z^C])\varepsilon^J. \nn
\end{align}
We can make use of the fact that the potential is
\begin{equation}
V = \frac{2}{3} \Tr ([Z^C , Z^D ; \bz_B ],[\bz_C , \bz_D ; Z^B]) - \frac{1}{3} \Tr ([Z^B , Z^D ; \bz_B ],[\bz_F , \bz_D ; Z^F])
\end{equation}
to write (c) as
\begin{align}
(c) = &-2 \delta^{IJ} (V-V_1)\varepsilon^J \nn \\
&- \G^{EF(IJ)}_{AB}\Tr ( [Z^C, Z^B ; \bz_C ], [\bz_E , \bz_F ; Z^A ])\varepsilon^J \nn \\
&- \G^{EF(IJ)}_{AB}\Tr ( [Z^A , Z^B ; \bz_E ],[\bz_C , \bz_F ; Z^C])\varepsilon^J \\
&+ \G^{EF(IJ)}_{AB}\Tr ( [Z^A, Z^B ; \bz_C ],[\bz_E, \bz_F ; Z^C])\varepsilon^J, \nn
\end{align}
where
\begin{equation}
V_1 = \frac{2}{3} \Tr([Z^C, Z^D ; \bz_B ],[\bz_C , \bz_D ; Z^B]) - \frac{4}{3} \Tr ([Z^C, Z^B ; \bz_C ],[\bz_E , \bz_B ; Z^E])
\end{equation}
\subsection{$\delta J^0$}
We can combine (a), (b) and (c) terms 
\begin{align}
\delta J^{0,I} =  &-2\delta^{IJ} T^0_\mu \gamma^\mu \varepsilon^J  + 2\delta^{IJ} V_1 \gamma^0 \varepsilon^J \nn \\
&+2 \delta^{IJ} ( \Tr (D_i \bz_B , [Z^D , Z^B ; \bz_D] ) -  \Tr (D_i Z^B , [\bz_D , \bz_B ; Z^D] ) \varepsilon^{ij}  \gamma^j \varepsilon^J \nn \\
&- \G^{C[IJ]}_{B}\Tr ( D_i Z^B , D_j \bz_C  )\varepsilon^{ij} \gamma^0 \varepsilon^J \nn \\
&- \G^{C[IJ]}_{B}( \Tr (D_0 \bz_A , [Z^B, Z^A ; \bz_C ] ) + \Tr ( D_0 Z^A, [\bz_C , \bz_A ; Z^B] )) \varepsilon^J \nn \\
&+ \G^{CD(IJ)}_{AB}( \Tr ( D^i \bz^B , [\bz_C , \bz_D ; Z^A] ) - \Tr ( D_i \bz_D [Z^A , Z^B ; \bz_C ] ))  \varepsilon^{ij} \gamma^j \varepsilon^J \nn \\
&- \G^{EF(IJ)}_{AB}\Tr ( [Z^C, Z^B ; \bz_C ], [\bz_E , \bz_F ; Z^A ])\varepsilon^J \nn \\
&- \G^{EF(IJ)}_{AB}\Tr ( [Z^A , Z^B ; \bz_E ],[\bz_C , \bz_F ; Z^C])\varepsilon^J\nn  \\
&+ \G^{EF(IJ)}_{AB}\Tr ( [Z^A, Z^B ; \bz_C ],[\bz_E, \bz_F ; Z^C])\varepsilon^J, \nn
\end{align}
where we have used
\begin{align}
T_{00} &= \Tr (D_0 Z^B , D_0 \bz_B ) + \Tr (  D_i Z^B , D^i \bz_B ) + V ; \\
T_{0i} &= \Tr ( D_0 Z^B , D_i \bz_B  ) + \Tr (  D_i Z^B , D_0 \bz_B ). 
\end{align}
\section{Potential}
In this appendix we show the equivalence of the Bagger-Lambert and ABJM potential. The Bagger-Lambert potential is given by
\begin{align}
V = \frac{2}{3} \Tr (\Upsilon^{CD}_B , {\bar{\Upsilon}}^B_{CD})
\end{align}
where
\begin{align}
\Upsilon^{CD}_B = [Z^C, Z^D;\bz_{B} ] - \frac{1}{2} \delta^C_B [Z^E, Z^D; \bz_E ] + \frac{1}{2}\delta^D_B [Z^E, Z^C; \bz_E ].
\end{align}
We can define the inner product as
\begin{align}
\Tr (X,Y) = \tr (X^\dagger Y)
\end{align}
where $\dagger$ denotes the transpose conjugate and $\tr$ denotes the ordinary matrix trace. Thus
\begin{align}
(\Upsilon^{CD}_B)^\dagger &= [Z^{D\dagger} , Z^{C\dagger} ; \bz_B^\dagger ] - \frac{1}{2} \delta^C_B [Z^{E \dagger} , Z^{D \dagger}, \bz^\dagger_{E}] + \frac{1}{2} \delta^D_B [Z^{E \dagger} , Z^{C\dagger} ; \bz^\dagger_{E}] \\
{\bar{\Upsilon}}^B_{CD} &= [\bz_C , \bz_D ; Z^B] - \frac{1}{2} \delta^B_C [\bz_E , \bz_D ; Z^E ] + \frac{1}{2} \delta^B_D [\bz_E , \bz_C ; Z^E ].
\end{align}
Making use of the above information one finds that
\begin{align}
V &= \frac{2}{3} \Tr (\Upsilon^{CD}_B , {\bar{\Upsilon}}^B_{CD}) \\
&=\frac{2}{3} \tr ((\Upsilon^{CD}_B)^\dagger {\bar{\Upsilon}}^B_{CD}) \\
&= \frac{2}{3} \tr \left([Z^{D \dagger} , Z^{C \dagger} ; \bz^\dagger_B ][\bz_C , \bz_D ; Z^B ] + \frac{1}{2} [Z^{E \dagger} , Z^{C \dagger} ; \bz^\dagger_E ][\bz_B , \bz_C ; Z^B ] \right). \label{bn}
\end{align}
For the particular choice
\begin{align}
[X, Y; Z] = \lambda (X Z^\dagger Y - Y Z^\dagger X) \label{vb}
\end{align}
it was shown by Bagger and Lambert that the $N=6$ ABJM potential is recovered. Inserting \eqref{vb} into \eqref{bn} one finds
\begin{align}
V = \lambda^2 \tr  ( &- \frac{1}{3} Z^{E \dagger} \bz_E Z^{C \dagger} \bz_C Z^{B \dagger} \bz_B - \frac{1}{3} Z^{C \dagger} \bz_E Z^{E \dagger} \bz_B Z^{B \dagger} \bz_C \nn \\
&- \frac{4}{3} Z^{D \dagger} \bz_B Z^{C \dagger} \bz_D Z^{B \dagger} \bz_C + 2 Z^{D\dagger} \bz_B Z^{C \dagger} \bz_{C} Z^{B \dagger} \bz_D ).
\end{align}
Comparing with
\begin{align}
V = \frac{4 \pi^2}{k^2}\tr (&- \frac{1}{3} X^A X_A X^B X_B X^C X_C - \frac{1}{3} X_A X^A X_B X^B X_C X^C  \nn\\
 &- \frac{4}{3} X_A X^B X_C X^A X_B X^C + 2 X^A X_B X^B X_A X^C X_C )
\end{align}
we see that the two expressions are equivalent given the redefinitions $Z^{A \dagger} \rightarrow X^A$ and $\bz_A \rightarrow X_A$, as well as $\lambda = 2 \pi/k$.

\end{appendix}

\bibliographystyle{abbrv}
\bibliography{main}

\end{document}